\documentstyle[11pt,titlepage]{article}  
\titlepage  
 
\begin{document} 
\begin{center} 
{\Large \bf Is Barbero's Hamiltonian formulation a Gauge Theory of Lorentzian
Gravity?}
\vskip 1cm 
{\large Joseph Samuel} \\ 
Raman Research Institute\\ 
Bangalore 560 080, INDIA.\\ 
\end{center} 
\vskip 2 cm 
\vskip 5 cm 
email:sam@rri.ernet.in 
\newpage  
\section*{Abstract}
This letter is a critique of Barbero's  constrained
Hamiltonian formulation of General Relativity on which
current work in Loop Quantum Gravity is based. While we do not
dispute the correctness of Barbero's formulation of general relativity,
we offer some criticisms of an aesthetic nature. We point out that unlike
Ashtekar's complex $SU(2)$ connection,  Barbero's real $SO(3)$ 
connection does not admit an
interpretation as a space--time gauge field.
We show that if one tries to interpret Barbero's real $SO(3)$ connection
as a space--time gauge field, the theory is not diffeomorphism invariant.
We conclude
that Barbero's formulation is not a gauge theory of gravity in the sense
that  Ashtekar's Hamiltonian formulation is. The advantages of Barbero's
real connection formulation have been bought at the price of giving
up the description of gravity as a gauge field.

\newpage
\section*{Introduction}
In the eighties, Ashtekar\cite{Ash} introduced  complex ``new variables''
on the phase space of General Relativity, which greatly 
simplified the form of the constraints. These variables
are a (densitised) soldering form and a complex $SU(2)$ 
connection. Ashtekar's motivation was to formulate General
Relativity in a manner similar to Yang-Mills fields,
so that ideas and techniques used in quantising gauge theories
could be imported into  Quantum Gravity. This motivation
was an attractive one, since the ``gauge'' description of Nature
is a unifying idea that seems to permeate diverse branches of physics.
Ashtekar suggested the use of a gauge field (``the connection representation'')
as the basic configuration variable in canonical gravity instead of the more 
traditional ``metric representation''. 
This area has been an active one \cite{gang}.
Witten's paper \cite{Witten} on the solution of $2+1$ gravity 
was partly inspired by Ashtekar's suggestion.
In 2+1 gravity the connection representation \cite{Witten,Achucarro,carlip}
has proved extremely useful and yields a considerably
simpler formulation of quantum gravity 
than the metrical description.
In both $2+1$ and in $3+1$ dimensions, the connection variable
is the pull-back of a space--time connection to a spatial slice ${\cal S}$.
However, in 3+1 dimensions, progress has been hampered
by the `reality conditions', which are necessary because
of the use of complex variables on the real phase space of General
Relativity.

In 1994, Barbero wrote a very influential paper \cite{Barbero}
pointing out that a small modification of Ashtekar's original
canonical transformation leads to a new Hamiltonian formulation of 
General Relativity in which the basic variables are 
\footnote{
These real variables were earlier considered by Ashtekar and discarded
in favour of the complex (``new'') variables of \cite{Ash} 
because the latter simplified the constraints.}
a {\it real} $SU(2)$ connection 
\footnote{
Or equivalently $SO(3)$; we will not make this relatively fine distinction 
in this paper.}
 and a {\it real} densitised triad.
The form of the Hamiltonian constraint in Barbero's formulation
is not as simple as in Ashtekar's original formulation.
But this is a small price to pay.
There are significant advantages in using real 
variables on the phase space of General Relativity,
because the `reality conditions' which had to be
imposed in Ashtekar's original formulation, are no
longer necessary. As a result, Barbero's Hamiltonian
formulation (BHF) has gained wide acceptance and is currently 
the basis of Loop Quantum Gravity. A lot of work \cite{work}
has 
been done on the space of real $SU(2)$ connections on manifolds.
Since $SU(2)$ is a compact group it has an invariant (Haar)
measure, and one is able to achieve 
a high degree of mathematical control over the space
of connections. 

It was pointed out by Immirzi \cite{Giorgio}
that Barbero's canonical transformation could be 
slightly generalised: a one parameter family of canonical
transformations were possible and all of them led to 
a Hamiltonian formulation based on a real $SU(2)$ connection.
This free parameter $\beta$ is known as the ``Immirzi parameter'',
and does not appear to be fixed by any theoretical considerations.

There are several puzzling features about Barbero's Hamiltonian
formulation. Viewed as a gauge theory (and this {\it was} Ashtekar's
original motivation), the gauge group of General Relativity
is certainly non-compact. Depending on the approach, one
might either believe the gauge group to be the Lorentz group
\cite{Utiyama} or the Poincare group \cite{Kibble}. It is 
generally agreed that the gauge group of General Relativity
must be non--compact. 
How then
is it possible to formulate GR as  gauge theory of a compact
group, as Barbero seems to?  
Of course, it {\it is} possible to formulate GR as a gauge theory
of a complex SU(2) group as in the original 
Ashtekar formulation. Over the complex numbers, there
is no distinction between compact and non--compact 
gauge groups. The puzzle is that Barbero's Hamiltonian 
formulation  uses a gauge group which is 
both  {\it real and compact}. Since BHF was derived by a canonical
transformation from a diffeomorphism invariant theory, it must
of course be diffeomorphism invariant. However,
there does not presently
exist any manifestly covariant Lagrangian formulation
of General Relativity as a gauge theory in which the
gauge group is real and compact. How does one understand this?

Holst\cite{Holst} has given a Lagrangian formulation 
equivalent to General Relativity and shown that on Legendre
transformation, it results in BHF. A curious feature of Holst's 
derivation is that he starts with
a gauge theoretic formulation with a non--compact $SO(3,1)$ gauge group. Yet
he is able to make contact with Barbero's formulation, which is based
on an $SO(3)$ gauge group. It is not clear {\it a priori} how the reduction
in the gauge group takes place.

Another puzzling feature of Barbero's Hamiltonian formulation
is that there are many of them. For every 
nonzero, real value of the Immirzi parameter $\beta$ there is 
a Hamiltonian formulation of General Relativity with 
as much claim to validity as Barbero's original formulation.
It is not easy to understand the origin of the `Immirzi
ambiguity'. It first appears as a parameter in a canonical
transformation that one performs on the phase space of GR.
However, instead of dissappearing from the final results
of the theory (as one might have hoped), it appears in the
spectrum of operators and also in the final expression
for Black Hole Entropy as calculated using Loop Quantum Gravity \cite{Krasnov}.
If the Immirzi parameter is indeed present in Quantum Gravity,
it would seem that a new fundamental constant (which can only
be fixed by experiment) has entered into physics. Such quantization
ambiguities do occur in other areas of  physics (e.g $\theta$ vacua),
but they are well understood to arise from a multiply connected
configuration space. The Immirzi ambiguity does not
appear to be of topological origin and does require understanding.

We emphasize that we do not question the correctness of BHF as a 
Hamiltonian formulation or its equivalence of BHF 
to General Relativity. This equivalence is assured since BHF was
derived from the ADM formulation of GR by a canonical transformation.
The aim of this paper is to understand the 
basis of BHF as a reformulation of General Relativity.
Is BHF a gauge theory of gravity? 
How does it happen that the gauge group is both real and compact?
How does the diffeomorphism invariance of BHF jive with the absence of
any manifestly covariant Lagrangian for GR as a real $SO(3)$ gauge
theory. 
In this paper we will answer these questions and thereby clarify some
aspects of Barbero's formulation.

Barbero's Hamiltonian formulation can be derived by making a 
canonical transformation starting from a constrained Hamiltonian
Formulation (CHF) due to Ashtekar-
the extended phase space construction (EPS). We note in 
passing that the EPS was an intermediate step in Ashtekar's
original derivation of the new variables. The EPS consists
of the following ingredients: the basic variables are
$({\tilde E}^a_i,K_a^i)$, which form a canonically conjugate
pair. The space--time meaning of these variables 
is that ${\tilde E}^a_i$ is a densitised triad on a spatial slice
${\cal S}$ and $K_a^i$ is the extrinsic curvature tensor of ${\cal S}$
with one index converted into a triad index. The three metric tensor
$q_{ab}$ of ${\cal S}$ is a derived object, which can be constructed 
from ${\tilde E}^a_i$ as explained in \cite{Ashtate}. Similarly,
the extrinsic curvature of ${\cal S}$ can also be expressed in terms of the 
basic fields. The constraints of the theory are
\begin{eqnarray*}
\epsilon_{ijk} K^{j}_{a} \tilde{E}^{ak} &\approx& 0 \label{SO(3)}\\
D_{a} [\tilde{E}^{a}_{k} K^{k}_{b} - \delta^{a}_{b} \tilde{E}^{c}_{k}
K^{k}_{c}] &\approx& 0\\
{\sqrt{q}} R + \frac{2}{\sqrt{q}} \tilde{E}^{[a}_{i}
\tilde{E}^{b]}_{j} K^{i}_{a} K^{j}_{b} &\approx& 0,
\end{eqnarray*}
where $D_{a}$ is the covariant derivative associated with
$q_{ab}$ and $R$, its scalar curvature. The Hamiltonian is a combination
of constraints. (We assume throughout this paper that space is closed
so that we can drop spatial boundary terms.)
The EPS  formulation is \cite{lag} strongly
diffeomorphism invariant (mod $SO(3)$ gauge) with the above
space--time interpretation for the basic variables $({\tilde E}^a_i,K_a^i)$.
We recall \cite{lag} that a Hamiltonian theory is strongly diffeomorphism
invariant (SDI) if a) there are constraints  whose brackets
reflect the Lie algebra of the diffeomorphism group. b) the basic
variables transform as is expected of them from their space--time
interpretation. For this criterion to be applied, one must
first declare the space--time interpretation of the basic variables
in the theory.

BHF differs from the EPS only by a canonical transformation. The basic
variables in BHF are $({\tilde E}^a_i,A_a^i)$, where 
${\tilde E}^a_i$ is the same as before and 
\begin{equation}
A_a^i:=\Gamma^i_a +\beta K^i_a,
\label{defBC}
\end{equation}
where $\beta$ is the Immirzi parameter and $\Gamma_a^i$
are the triad spin coefficients. The constraints of the theory are now
\begin{eqnarray}
{\cal D}_{a} \tilde{E}^{a}_{i} &\approx& 0\label{BHFgauss}\\
\tilde{E}^{b}_{i} F^{i}_{ab} &\approx& 0\label{BHFvector}\\
\epsilon^{ijk} \tilde{E}^{a}_{i} \tilde{E}^{b}_{j} F_{abk} -2\frac{(1+\beta^2)}{\beta^2} \tilde{E}^{a}_{[i} 
\tilde{E}^{b}_{j]} (A^{i}_{a} - \Gamma^{i}_{a}) 
(A^{j}_{b} - \Gamma^{j}_{b}) &\approx& 0\label{BHFscalar},
\end{eqnarray}
where ${\cal D}$ is the covariant derivative associated with the
Barbero connection (\ref{defBC}).
Note that the Scalar constraint (\ref{BHFscalar}) explicitly contains 
the Immirzi parameter. 
If one now endows $A^i_a$ with a space--time interpretation coming
from (\ref{defBC}) as a linear combination of the triad spin coefficients
of ${\cal S}$ and the extrinsic curvature of ${\cal S}$, it is clear that
the new formulation is {\it still} SDI (mod $SO(3)$ gauge),
since the starting point was.

The original Ashtekar variables were derived by 
making the canonical transformation (\ref{defBC}) with $\beta = i$ (or
equivalently $-i$). This choice is distinguished in several ways.\\
{\it Why $\beta = i$ is special:}  
\begin{enumerate}
\item {\it Simplification of the Hamiltonian constraint:} 
The expression (\ref{BHFscalar}) for the
Hamiltonian constraint in BHF  contains a 
messy term multiplied by $(1+\beta^2)$.
For $\beta = \pm i$ this term disappears and we get a simplified
form for the Hamiltonian constraint.
\item {\it Full local Lorentz invariance:}
As emphasized by Immirzi \cite{Giorgio} the time gauge is
not needed to derive the original Ashtekar variables. They can 
be derived from a manifestly covariant action \cite{Selfdual} 
{\it maintaining full
local Lorentz invariance}. It 
follows that the theory is manifestly SDI, (i.e., SDI and not just
modulo gauge).
\item {\it Space--time interpretation for the connection:}
The Ashtekar connection (as defined by (\ref{defBC}) with $\beta = i$)
can also be interpreted as a space--time connection.
More precisely, the Ashtekar connection
is the pull back to ${\cal S}$ of a space--time connection 1-form. This
is evident because Ashtekar's formulation can be derived from a manifestly
covariant Lagrangian \cite{Selfdual} in which one of the basic fields is an $SL(2,{\mbox{$I\!\!\!\!C$}})$
space--time connection. The Ashtekar connection then
appears as the pull-back of the space--time $SL(2,{\mbox{$I\!\!\!\!C$}})$ connection
to ${\cal S}$. Thus we would regard Ashtekar's Hamiltonian Formulation
(AHF) as a gauge theoretic
reformulation of General Relativity.
\end{enumerate}
For values of $\beta$ other than $\pm i$ these  three properties do not
obtain. 
\begin{enumerate}
\item
As Barbero \cite{Barbero} points out, the
form of the Hamiltonian constraint is more complicated for $\beta$ real, 
but one can choose to accept it. 
\item It appears that one must choose the ``time gauge" in order
to arrive at Barbero's formulation. In this formulation the full
local Lorentz invariance of the Ashtekar formulation is lost.
One can also choose to live with this, since this is
only a choice of gauge.
\item
The Barbero connection does not have a space--time interpretation
as the pull back of a space--time connection 1-form to ${\cal S}$.
In the case $\beta = i$, one had the option of interpreting the $A^{i}_{a}$ 
defined by (\ref{defBC}) as a space--time connection. If one attempts to 
do this for real $\beta$, one finds that the theory violates Strong
Diffeomorphism Invariance. Under 
diffeomorphisms tangential to ${\cal S}$, the $A^i_a$ defined by (\ref{defBC})
does transform like a connection
1-form. However, for diffeomorphisms that are normal to ${\cal S}$, the 
canonical transformation properties of the $A^i_a$ do not reflect its 
proposed
space--time interpretation as the pullback of a space--time 
connection 1--form. This point is explained and proved below.
\end{enumerate}

{\it Claim: 
Barbero's connection cannot be interpreted as a space--time connection}\\
Proof: Consider a solution $({\cal M},g)$ of Einstein's equations and
a loop $\gamma$ in ${\cal M}$. If a connection transforms correctly
under space--time diffeomorphisms (i.e, as a 1-form), 
the trace of the Holonomy of the  
connection along $\gamma$ should depend only on the loop $\gamma$ and should
not depend on the slicing of $\cal M$. It is easy to construct an example
of an empty space solution to Einstein's equations in which the
Holonomy of the Barbero Connection 
changes from being trivial for one slice containing $\gamma$  to 
non --trivial for another slice containing $\gamma$.
Consider flat space--time $({\cal M},\eta)$ with standard Minkowski
coordinates $(t,x,y,z)$  and a loop $\gamma$ which 
is a circle of radius $R$ described by  $t=\sqrt{1+R^2}, z= 0,x=R\cos \theta,
y=R \sin \theta, 0\le \theta \le 2\pi$.
The flat slice ${\cal S}_1$ defined by $t=\sqrt{1+R^2}$ contains $\gamma$. 
Since both the intrinsic and extrinsic curvature of ${\cal S}_1$ vanish,
the holonomy of the Barbero connection is the identity and its 
trace is $3$. However, the same loop is also contained in the  
hyperbolic
slice, ${\cal S}_2$ defined by  $t^2-x^2-y^2 = 1$. We now
compute the trace of the Holonomy $tr H(A)$ of the Barbero
connection along the same loop $\gamma$.

At the point $y^\mu=(t=1,\vec{x}=0)\in {\cal S}_2$  
the standard orthonormal frame ${\hat e}^I_{}\mu=\delta^I{}_\mu$ ($I=0,1,2,3$
is a frame index and $\mu=0,1,2,3$ is a space-time index) has the property that
${\hat e}^0{}_\mu$ is normal to ${\cal S}_2$. Let us move this frame to
other points  $x^\mu\in{\cal S}_2$  by the Lorentz transformation
\begin{equation}
e^I{}_\mu(x):=\Lambda_\mu{}^\nu(x) {\hat e}^I{}_\nu,
\label{move}
\end{equation}
where
\begin{equation}
\Lambda_\mu{}^\nu:=\delta_\mu{}^\nu+(1-x.y)^{-1}(x_\mu+y_\mu)(x^\nu+y^\nu)-2x_\mu
y^\nu.
\label{Lambda}
\end{equation}
 $\Lambda_\mu{}^\nu$ has the property that $x_\mu=\Lambda_\mu{}^\nu y_\nu$,
 which ensures that $e^0{}_\mu$ is normal to ${\cal S}_2$ everywhere. (Our
 frame satisfies the `time gauge'.) The space-time connection 1-form is given by $A_\alpha{}^{IJ}$
 \begin{equation}
 A_\alpha{}^{IJ}:=e^I_\mu \partial_\alpha e^{\mu J}
 \label{condef}
 \end{equation}
 ($A^{IJ}$ with {\it two} internal indices is the space-time connection and should
 not be confused with the Barbero connection $A^i$, 
 which has one internal index.) The contraction $A^{IJ}:=t^\alpha A_\alpha$
 of the connection one-form with the tangent vector 
 $t^\alpha$ to the curve $\gamma$ is
 easily worked out. The non-vanishing components are
 $A^{01}=-A^{10}=-y,A^{02}=-A^{20}=x,A^{12}=-A^{21}=1-t$
 Constructing Barbero's  connection by the formula
 \begin{equation}
 A^i=1/2 \epsilon^{ijk} A^{jk}+\beta A^{0i},
 \label{Barberocon}
 \end{equation}
 we find that $A^1=-\beta y, A^2=\beta x, A^3= 1-t$.
 An elementary calculation then yields the result
 \begin{equation}
 {\rm Tr} H(A)=1+2 \cos (2\pi \sqrt{1+R^2(1+\beta^2)})
 \label{TrHA}
 \end{equation}
 
   Notice that the trace of the Holonomy of
the Barbero connection along the same loop $\gamma$,  
depends on $\beta$ and (except for the special values $\beta=\pm i$) 
is  not the equal to 
$3$. Briefly, the trace of the holonomy of the Barbero Connection 
along a loop $\gamma$ is not just a property of the loop
but also depends on the slicing.

What this means is that it is impossible to attach a gauge
theoretic space--time
interpretation to Barbero's connection. As a {\it spatial} connection,
Barbero's connection is certainly well defined and, in fact, 
transforms correctly under
diffeomorphisms that preserve the spatial slice. But unlike Ashtekar's
connection, Barbero's connection does not admit an interpretation
as a space--time gauge field. Such an interpretation is not
consistent with the Poisson bracket relations between $A_a^i$ and
the Scalar constraint. It appears that the ``extra'' term proportional
to $1+\beta^2$ ruins this bracket. Thus  BHF is not a gauge theory
of gravitation.

This observation removes most of the puzzles we had raised
at the beginning of this letter.  BHF is not a gauge theory
of gravitation and so one cannot identify Barbero's $SO(3)$ 
with the gauge group of gravity. There is no conflict between
Barbero's Hamiltonian formulation and our expectation that the
gauge group of gravity must be noncompact.

This observation also removes the puzzle of how Holst is able
to derive a compact gauge group from a non compact one. The variable
defined as a connection variable by Holst \cite{Holst} is not the
pullback of a space--time connection. Rather, certain components
of the space--time connection are defined to be components of a new
$SO(3)$ connection. This is Barbero's connection and it has no
space--time significance.
One sometimes
sees it implied \cite{Giorgio,Holst} that the reduction in the gauge
group from $SO(3,1)$ to $SO(3)$ takes place because of our choice of
the ``time gauge''. It is indeed true that once we make 
this gauge choice\footnote{
Holst's orginal derivation of BHF from his covariant Lagrangian
contained a small logical gap: he used the time gauge fixed
action to perform the Legendre transformation. This gap
has since been filled in papers by S. Alexandrov \cite{Sergei} 
and Nuno Barros e Sa \cite{Sa}
},
our freedom to make additional gauge transformations is curtailed from
$SO(3,1)$ to $SO(3)$. However this does {\it not} mean that the gauge
group has been reduced. The pullback of the connection to a spatial slice
is still an $SO(3,1)$ connection, in spite of our gauge choice.

While this point is elementary, it is worth making in detail 
and we digress briefly to do so. The central point here 
is a geometrical notion: the holonomy group of a connection 
\cite{Cartan}.
Given a $G$ connection $A$ on a manifold $\cal S$, we define its holonomy
group to be the collection $H_p(A) = \{H_{\gamma_p}(A)\}$ 
of elements of
$G$ which arise as holonomies of $A$ along closed loops $\gamma_p$ 
based at $p$. 
For example, if the connection is pure gauge, the holonomy group is trivial. 
It is easily seen that the holonomy groups based at $p$ and 
$p^\prime\in
{\cal S}$ are related by conjugation and therefore isomorphic. 
Further, under a local  
gauge transformation, 
the holonomy group of $A$ transforms by conjugation
\begin{equation}
H_p(A) \rightarrow g(p) H_p(A) g^{-1}(p).
\end{equation}
where $g(p)\in G$. 
We say that a connection $A$ is {\it reducible} if its holonomy group
is a proper subgroup of $G$. Clearly, {\it the reducibility of a connection
is a gauge invariant notion and is independent of gauge choice}. A general
$SO(3,1)$ connection is not reducible to $SO(3)$.

One may of course, simply abandon the ``gauge interpretation''
of the theory and view the Barbero connection as a purely spatial
connection which one uses for technical reasons to produce quantum 
states which are functionals of the extrinsic curvature.
But then one must be aware that one has given up the gauge interpretation.
We argue here that there are strong aesthetic reasons for retaining
the gauge interpretation. One of the principal motivations of the Ashtekar
program was the gauge description of gravity. This appears to be a unifying
thread across the different forces of nature: they are all described by
a space-time gauge field.
The view we would like to offer
here is that this was an important motivation of the original Ashtekar
program and should be retained. 
From this point of view, the Immirzi
parameter is not a free parameter but must be fixed to the special
value $i$. This is one possible resolution of the ``Immirzi Ambiguity''.
Similar views have been expressed (see footnote 3 of \cite{Sergei})
by Alexandrov \cite{Sergei}.

If one gives up the gauge interpretation of gravity, the Immirzi
parameter appears not to be fixed by theory. This would not be a problem
if the parameter dissappeared 
\footnote{as in Alexandrov's path integral quantisation \cite{Sergei}.}
from all physical predictions 
of the theory.
However, this is not the case: the Immirzi parameter does appear
in the calculated value of Black Hole entropy in Loop Quantum Gravity. 
This phenomenon does not appear to be well understood 
(unlike the $\theta$ vacua of QCD). Rovelli 
and Thiemann \cite{thiem} offer a finite dimensional example of the Immirzi
ambiguity. We do not find this example convincing: 
their original system does not suffer from any ambiguity.
The ambiguity is introduced `by hand' by changing the original configuration
space in a $\beta$ dependent manner. Thus their ``quantisation prodedure''
does not quantise the original system at all, but quantises a one parameter
family of distinct systems \cite{Imm}.

If one accepts the idea that canonical transformations made by the
theorist can introduce parameters into the physical predictions of 
the quantum theory, one
does lose predictive power. Admittedly, this loss is very small in
the case of the Immirzi ambiguity. The ambiguity is to the extent of 
a single parameter which can be fixed by comparing a single prediction
of the theory with an independent calculation (say the Black Hole entropy)
or (thinking wishfully) an experiment. 
However, we would argue that in a field theory like General Relativity, 
there are infinitely
many degrees of freedom (two at each spatial point) 
and one could, in principle, contemplate making separate 
canonical transformations in each of them, thereby introducing
an enormous ambiguity into the theory. In the absence of any 
internal criterion to curb this ambiguity, the physical predictions
of the theory may depend on an infinite number of parameters.
Such a theory would have no predictive power.

If one wishes to describe Euclidean Gravity rather than Lorentzian Gravity,
the gauge group is automatically compact and the Ashtekar variables
are real. Barbero's connection (with $\beta=1$) agrees with Ashtekar's
and one does have a real formulation of Euclidean gravity as a gauge
theory. However, for $\beta$ not equal to unity, the same problem 
arises: Barbero's connection is not a space--time connection and
if one attempts to interpret it as such, one violates diffeomorphism
invariance. It appears then that the values $\beta=1$ (for Euclidean)
and $\beta=i$ (for Lorentzian Gravity) are very special. One could hope
to relate these two by a Wick rotation as discussed in \cite{wick}

In this paper we argue strongly for maintaining the gauge
aspect of gravity in the approach to quantum gravity, even though
the gauge group is non--compact and therefore not as tractable as
say, $SU(2)$.
It does not appear to us a strong argument to say that we study
compact gauge groups because we do not know how to deal with non--compact
gauge groups with mathematical rigour. The non--compactness
of the gauge group appears to us an essentially physical feature of General
Relativity,  which is closely related to the Minkowskian signature
of the space--time metric and light cones. Phenomena like infinite
red shift, seen in Black Hole physics need non--compact groups for a gauge
theoretical description. We would suggest that one must learn to deal
with non--compact gauge groups. One could return either to the original
Ashtekar variables with a complex connection or to the real Palatini
tetrad formulation, with non--compact gauge group $SO(3,1)$. 
In our opinion these two problems 
-- non--compact gauge groups and complex gauge groups-
 are the same problem in different guises.
Evading both problems simultaneously is
impossible, if 
one is interested in dealing
with the physical effects of Lorentzian General Relativity as a gauge theory.

{\it Acknowledgement:} It is a pleasure to thank Richard Epp,
B.R. Iyer, Sukanya Sinha and Madhavan Varadarajan for 
extended discussions and Ramesh Anishetty, Ghanashyam Date, N.D. Haridass,
T.R. Govindrajan, Romesh Kaul and H.S Sharatchandra
for their critical comments on this work.


\begin{thebibliography}{99}
\small
\setlength{\itemsep}{-0.9\parsep}


\bibitem{Ash}
Ashtekar A (1986) {\it Phys. Rev. Lett.} {\bf 57} 2244;
Ashtekar A (1987) {\it Phys. Rev.} {\bf D36} 1587;\\
see also \cite{Ashtate}.
\bibitem{gang} For exhaustive references in this area,
see the review article by Gaul, M and Rovelli C, gr-qc/99100 79.
\bibitem{Witten}
Witten E (1998) {\it Nucl. Phys.} {\bf B311} 46.
\bibitem{Achucarro}
Achucarro A and Townsend P K (1986) {\it Phys. Lett.} {\bf B180} 89.
\bibitem{carlip}
Carlip S (1998) {\it Quantum Gravity in 2+1 Dimensions}, CUP, Cambridge.
\bibitem{Barbero} Barbero, J F Phys. Rev. D., {\bf 51}, 5507 (1995),
gr-qc/9410014.
\bibitem{work}
see Rovelli, C. {\it Loop Quantum Gravity}, Living Reviews gr-qc 9710008.
\bibitem{Giorgio}
Immirzi, G (1997) Class. Quant. Gravity {\bf 14}, L177-181.
gr-qc/9511026.
\bibitem{Utiyama}
Utiyama R, 1956 {\it Phys. Rev} {\bf 101} 1597.
\bibitem{Kibble}
Kibble T W B, 1961 {\it Journ. Math. Phys.} {\bf 2} 212.
\bibitem{Holst} Holst S (1996) {\it Phys. Rev.} {\bf D53}, 5966,
gr-qc/9511026.
\bibitem{Krasnov}
Ashtekar A, Baez J, Corichi A and Krasnov K (1998) {\it
Phys. Rev. Lett.} {\bf 80}, 904.\\
Kaul, R K and Majumdar P (2000), {\it Phys. Rev. Lett} (to appear)
gr-qc/0002040.
\bibitem{Ashtate}
 Ashtekar A (notes prepared in collaboration with R. Tate) (1991) 
{\it Lectures on Non-perturbative
canonical gravity}, World Scientific, Singapore.
\bibitem{lag}
Samuel J, (2000) 
{\it Canonical Gravity, Diffeomorphisms and Objective Histories}
preprint.
\bibitem{Selfdual} Jacobson T and Smolin L (1988) {\it Class. and Quant. Grav.}
 {\bf 5} 583;
 Samuel J (1987) {\it Pramana-J. Phys.} {\bf 28} L429.
\bibitem{Cartan}
Cartan E, (1986) `On Manifolds with an affine connection and the 
theory of General Relativity', English translation by 
Anne Magnon and Abhay Ashtekar, Bibliopolis, Naples.
\bibitem{Sergei}
Alexandrov S (2000) ``$SO(4,C)$- covariant Ashtekar-Barbero gravity and
the Immirzi parameter'', gr-qc/0005085
\bibitem{Sa}
Barros e Sa N (2000) {\it ``Hamiltonian Analysis of General Relativity with the
Immirzi Parameter''} gr-qc/0006013.
\bibitem{thiem}
Rovelli C and Thiemann T, {\it The Immirzi parameter in quantum general
relativity} gr-qc 9705059.
\bibitem{Imm}
Samuel J (2000) {\it ``Comment on the Immirzi parameter 
in Quantum General Relativity''}, submitted to Physical Review .
\bibitem{wick}
Ashtekar A (1996) {\it Phys. Rev.} {\bf D53} 2865.
\end{thebibliography}
\end{document}